\documentclass[aps,twocolumn,prl,amsmath,amssymb,showpacs,notitlepage,superscriptaddress]{revtex4-1}
\usepackage{graphicx}				% Use pdf, png, jpg, or eps§ with pdflatex; use eps in DVI mode
\usepackage{bm}							% TeX will automatically convert eps --> pdf in pdflatex		
\usepackage{amssymb}
\usepackage{amsmath}
\usepackage{color}
\usepackage{epstopdf}

\begin{document}

\title{Anomalous Hall effect driven by dipolar spin waves in uniform ferromagnets}

\date{\today }

\author{Kei Yamamoto}
\email{kei@sci.phys.kobe-u.ac.jp}
\affiliation{Department of Physics, Kobe University, 1-1 Rokkodai, Kobe 657-8501, Japan}
\affiliation{Institut f\"ur Physik, Johannes Gutenberg Universit\"at Mainz, D-55099 Mainz, Germany}
\author{Koji Sato}
\affiliation{WPI Advanced Institute for Materials Research, Tohoku University, Sendai 980-8577, Japan}
\author{Eiji Saitoh}
\affiliation{WPI Advanced Institute for Materials Research, Tohoku University, Sendai 980-8577, Japan}
\affiliation{Institute for Materials Research, Tohoku University, Sendai 980-8577, Japan}
\affiliation{Spin Quantum Rectification Project, ERATO, Japan Science and Technology Agency, Sendai 980-8577, Japan}
\affiliation{Advanced Science Research Center, Japan Atomic Energy Agency, Tokai 319-1195, Japan}
\author{Hiroshi Kohno}
\affiliation{Department of Physics, Nagoya University, Furo-cho, Chikusa-ku, Nagoya, 464-8602, Japan}

\begin{abstract}
A new type of anomalous Hall effect is shown to arise from the interaction of conduction electrons with dipolar spin waves in ferromagnets. 
 This effect exists even in homogeneous ferromagnets without relativistic spin-orbit coupling. 
 The leading contribution to the Hall conductivity is proportional to the chiral spin correlation 
of dynamical spin textures and is physically understood in terms of the skew scattering by dipolar magnons. 
\end{abstract}

\preprint{KOBE-TH-15-08}

\maketitle

The anomalous Hall effect (AHE) has been known for more than a century and provided various intriguing areas of researches both theoretically and experimentally \cite{Nagaosa2010}. The AHE are due to both the presence of the spontaneous magnetization and spin-orbit interaction, and thus no external magnetic field is needed as opposed to the ordinary Hall effect. 
 The contributions to the AHE are conventionally understood by the intrinsic and extrinsic mechanisms; the former is associated with the anomalous velocity deriving from non-trivial band curvature (Berry curvature), and the latter comes from spin-orbit scattering from impurities \cite{Karplus1954,*Smit1955,*Smit1958,*Luttinger1958,*Berger1970,*Berger1972}. 
 Distinctly from these mechanisms, another type of contribution to the AHE, known as the topological Hall effect (THE), was later proposed \cite{Ye1999, Bruno2004}. The THE arises from chiral magnetic textures such as Skyrmion configuration, which generate Berry phase for conduction electrons leading to Hall effect \cite{Nagaosa2013}. This Berry phase interpretation is appropriate in the regime of strong coupling between conduction electrons and magnetic textures. 
 On the other hand, the Hall conductivity in the weak coupling regime was shown to be proportional to the scalar spin chirality, given by $\bm{S}_{\bm{r}_1} \cdot ( \bm{S}_{\bm{r}_2} \times \bm{S}_{\bm{r}_3})$, which measures the correlation between the spin orientations and real-space positions \cite{Tatara2002}. The leading order contribution reads
\begin{equation}
\sigma _{xy} \propto \sum _{\bm{r}_1,\bm{r}_2,\bm{r}_3} \bm{S}_{\bm{r}_1} \cdot \left( \bm{S}_{\bm{r}_2} \times \bm{S}_{\bm{r}_3} \right) \left( \bm{a}\times \bm{b} \right) _z \mathcal{F}\left( a,b,c\right) , \label{eq:Tatara}
\end{equation} 
where $\bm{a}=\bm{r}_1 -\bm{r}_2 , \ \bm{b}=\bm{r}_2 - \bm{r}_3 ,\ \bm{c} = \bm{r}_3 -\bm{r}_1 $,  
$\mathcal{F}\left( a,b,c\right) $ is a symmetric function of $a=|{\bm a}|$, $b=|{\bm b}|$ and $c=|{\bm c}|$ 
related to electrons, 
and the summation is taken over all locations ${\bm r}_i$ of localized magnetic moments ${\bm S}_{{\bm r}_i}$. In order to have a non-vanishing contribution in this setting,
the scalar spin chirality must include a component that is odd under parity 
(i.e. $\bm{r}_i  \equiv (x_i, y_i, z_i)  \rightarrow  (\pm x_i, \mp y_i, z_i)$), 
which, in translationally and rotationally symmetric systems, requires $\bm{S}_{\bm{r}_1} \cdot ( \bm{S}_{\bm{r}_2} \times \bm{S}_{\bm{r}_3}) |_{\rm odd} \propto (\bm{a}\times \bm{b} )_z $.

Recently, Hall effects not only of electrons but also of various bosonic excitations, involving photons, magnons and phonons, have been studied \cite{Onoda2004,*Haldane2008,*Onose2010,*Qin2012,*Shindou2013}. Amongst them is the Hall effect of magnetostatic forward volume waves, or dipolar magnons, which have Berry curvature originating from the magnetic dipole-dipole interaction  \cite{Matsumoto2011}. 
 Since conduction electrons interact with magnons via the exchange interaction, it seems natural to expect that the electrons may also acquire some chirality in their orbital motion and exhibit a Hall effect.

 In this Letter, we approach this problem by a perturbative expansion in the exchange coupling
 between the dynamical magnetization and conduction electrons.
 We derive a spin-chirality formula for the Hall conductivity, which can be regarded as a generalization
 of Eq.~(\ref{eq:Tatara}) to spin texture fluctuations. It is found that the dipole-dipole interaction induces a %non-vanishing 
parity-odd component of {\it chiral spin correlation},
\begin{eqnarray}\label{eq:chi_1}
\hspace{-0.5cm}\left\langle \bm{S}_{\bm{r}_1}\cdot ( \bm{S}_{\bm{r}_2} \times \bm{S}_{\bm{r}_3} ) \right\rangle |_{\rm odd} &=& \left( \bm{a}\times \bm{b} \right) _z \left( \bm{a}\cdot \bm{b} \right) _t \tilde{F}( a) \tilde{F}(b) \nonumber\\
&& + {\rm cyclic \ perms \ in \ }(a,b,c) ,
\end{eqnarray}
 where $\langle \cdots \rangle $ denotes quantum/thermal averaging, 
the subscript $t$ indicates projection onto $x$-$y$ plane $(\bm{a}\cdot \bm{b})_t = a_x b_x + a_y b_y$, 
and $\tilde{F}(a)$ is a function related to magnons. 
 Because of the factor $({\bm a} \cdot {\bm b})_t$ in Eq.~(\ref{eq:chi_1}), 
there is no net 
%chiral spin correlation 
spin chirality in the sense that the spatial average of $\langle \bm{S}_{\bm{r}_1}\cdot (\bm{S}_{\bm{r}_2} \times \bm{S}_{\bm{r}_3} ) \rangle  (\bm{a}\times \bm{b})_z $ is zero. Nevertheless, the electron contribution, which corresponds to $(\bm{a} \times \bm{b})_z \mathcal{F}$ in Eq.~(\ref{eq:Tatara}), provides factors of $(\bm{a}\cdot \bm{b})_t$ in such a way to render the Hall conductivity finite, as we shall demonstrate below.

Let us first reformulate the theory of dipolar magnons in a form suitable for studying their interaction with electrons. 
 We consider a normally magnetized film of homogeneous ferromagnet. 
 The magnetization $\hat{\bm{m}}\left( \bm{r} \right) $ satisfies
$ \left\langle \hat{\bm{m}}\left( \bm{r} \right) \right\rangle _0 = \left( 0, 0, M_s \right)$ in the ground state, 
where $M_s$ is the saturation magnetization. 
 Let us introduce the Holstein-Primakoff field $a_{\bm{q}}$ by
\begin{equation}
 \hat{m}_+ \left( \bm{r} \right) 
 = \sqrt{\frac{-2\hbar \gamma M_s}{V}}\sum _{\bm{q}} a_{\bm{q}} e^{i\bm{q}\cdot \bm{r} }  \,, 
\end{equation}
where $\hat{m}_\pm = \hat{m}_x  \pm i \hat{m}_y$, $\gamma <0$ is the effective gyromagnetic ratio of electron, and $V$ is the volume of the sample.
 In the presence of both exchange and dipole-dipole interactions, the magnon Hamiltonian can be written as \cite{Stancil2009}
\begin{equation}\label{magnon Hamiltonian}
 H_{\rm d} = \sum _{\bm{q}} \hbar \omega _{\bm{q}} b^{\dagger }_{\bm{q}} b_{\bm{q}} , \  
\end{equation}
in terms of the diagonalizing operator, 
$ b_{\bm{q}} = A^+_{\bm{q}}e^{-i\phi _{\bm{q}}} a_{\bm{q}} +A^-_{\bm{q}}e^{i\phi _{\bm{q}}} a^{\dagger }_{-\bm{q}} $, and the eigenfrequency 
\begin{equation}
\omega _{\bm{q}} = \sqrt{\left( \omega _0 +\omega _M \lambda q^2 \right) \left( \omega _0 + \omega _M \lambda q^2 + \omega _M \sin ^2 \theta _{\bm{q}}\right) }\, , 
\end{equation}
where 
$ \bm{q} = q ( \sin \theta _{\bm{q}} \cos \phi _{\bm{q}} , \sin \theta _{\bm{q}}\sin \phi _{\bm{q}} ,\cos \theta _{\bm{q}} )$. 
 The Bogoliubov coefficients are given by 
\begin{equation}
 A^{\pm }_{\bm{q}} = \frac{\omega _{\bm{q}}\pm \left( \omega _0 + \omega _M \lambda q^2 \right) }{2\sqrt{\omega _{\bm{q}}\left( \omega _0 + \omega _M \lambda q^2 \right) }}\,.
\end{equation}
 If we neglect magnetocrystalline anisotropy, $\omega _0 = -\gamma H_{\rm ex}$ and $\omega _M = -\gamma M_s $, where $H_{\rm ex}$ is an external magnetic field, and $\lambda $ is the square of the exchange length, whose typical value is $\sim 10^{-16} [{\rm m}^2]$. 
 Note that $a_{\bm{q}}$ is not an eigen operator unless the dipole-dipole interaction is absent, $A^-_{\bm{q}}=0$. 
 To treat interactions between magnons and electrons, we shall apply quantum field theory techniques \cite{Abrikosov1975,Woolsey1970,*Danon2014}
and define thermal Green's functions
\begin{equation}
 {\cal D}_{AB}\left( 1,2 \right) = \frac{1}{2\hbar \gamma M_s} \left\langle {\rm T}_{\tau }\left[ \hat{m}_A \left( \tau _1 ,\bm{r}_1 \right) \hat{m}_B  \left( \tau _2 ,\bm{r}_2 \right) \right] \right\rangle\,, \label{eq:magnon_Green}
\end{equation}
where ${\rm T}_{\tau }$ indicates imaginary-time ordering, $A,B = \pm $, and 
$\hat{m}_A ( \tau ) = e^{H\tau } \hat{m}_A  e^{-H\tau }$. 
 A consequence of the dipole-dipole interaction is the spin-flip propagator
\begin{equation}\label{Dpm}
\mathcal{D}_{\pm \pm} = - \frac{k_B T}{V}\sum _{\ell ,\bm{q}}e^{\pm 2i\phi _{\bm{q}}}A^+_{\bm{q}}A^-_{\bm{q}}\mathcal{D}_{\bm q} \left( i\nu _{\ell } \right) e^{i ( \bm{q}\cdot \bm{r} - \nu _{\ell }\tau ) } \,,
\end{equation}
where $\nu _{\ell }$ is the Matsubara frequency and 
$\mathcal{D}_{\bm q} \left( i\nu _{\ell } \right) =  - 2\hbar \omega _{\bm{q}}/ (\nu _{\ell }^2 + \hbar ^2 \omega _{\bm{q}}^2 )$.
 We neglect the self-energy for simplicity. 
The phase factor $e^{\pm 2i\phi _{\bm{q}}}$ in Eq.~(\ref{Dpm})
is a manifestation of the coupling between the spin and orbital motion of magnons. 
 In particular, the spin-flip propagator (\ref{Dpm}) contains a parity-odd element, which will ultimately lead to electron Hall effect.  
 The spin-preserving part $\mathcal{D}_{\pm \mp} $ also receives a contribution from the dipole-dipole interaction, but it is parity even and does not essentially affect AHE.

 For the electron part of Hamiltonian, we consider a free electron gas with the {\it s-d} exchange coupling to $\hat {\bm m}$; 
\begin{eqnarray}
 H_{\rm s} &=& \sum _{\bm{k}} \xi _{\bm{k}} c^{\dagger }_{\bm{k}} c_{\bm{k} }  \,,\quad 
\xi _{\bm{k}} =\frac{\hbar ^2 k^2}{2m} - \mu \,,\\
 H_{\rm sd} &=&  - \frac{J }{N} \sum _{\bm{k},\bm{q}}\int d\bm{r} \frac{\hat{\bm{m}}\left( \bm{r}\right) }{\hbar \gamma } \cdot \left( c^{\dagger }_{\bm{k}}\bm{\sigma }c_{\bm{k}+\bm{q}}\right) e^{i\bm{q}\cdot \bm{r}}\,.
\end{eqnarray}
 Here, $c_{\bm{k}}^\dagger = (c_{\bm{k} \uparrow}^\dagger , c_{\bm{k} \downarrow}^\dagger)$ is the electron creation operator, 
$\mu$ is the chemical potential, $J$ the {\it s-d} exchange energy, $N$ the total number of localized magnetic moments in the sample and $\bm{\sigma }$ the triplet of the Pauli matrices. 
 Our goal is to compute the Hall conductivity $\sigma_{xy}$ based on the Hamiltonian $H= H_{\rm s}+ H_{\rm d}+ H_{\rm sd}$ and the Kubo formula
\begin{equation}\label{Kubo formula}
\sigma _{xy}\left( \omega \right) = \frac{V}{i\omega} \int d\tau \left\langle {\rm T}_{\tau }\left[ j_x \left( \tau \right)  j_y  \right] \right\rangle e^{i\omega_\lambda \tau }  \,,
\end{equation}
where $ {\bm j} = \frac{\hbar e}{mV}\sum _{\bm{k}} {\bm k} c^{\dagger }_{\bm{k} } c_{\bm{k} }$ is the charge-current density operator,
and the analytic continuation $i\omega_\lambda \rightarrow \omega +i 0 $ is understood. 
 The electron propagator is given by 
$  \mathcal{G}_{{\bm k} \sigma} \left( i\varepsilon_n \right) = \left\{ i\varepsilon_n - \xi_{\bm{k} \sigma} -\Sigma_{{\bm k} \sigma} \left( i\varepsilon_n \right) \right\} ^{-1}$ 
with $\xi_{\bm{k} \sigma} = \xi_{\bm{k}} +JS \sigma$, 
where $S = -M_s V / \hbar \gamma N $ is the magnitude of localized spins, 
$\sigma = \pm 1$ the spin projection, and $\Sigma $ is the self-energy. 
 We take $\Sigma_{{\bm k} \sigma} ( \varepsilon + i0 ) = -i \Gamma_{{\bm k} \sigma} ( \varepsilon )$, where $\Gamma$ is the  damping parameter, neglecting its real part.

  We assume $J$ is small and calculate $\sigma_{xy}$ perturbatively with respect to $J$. 
  As in Ref.~\cite{Tatara2002}, a non-trivial contribution shows up at the third order. 
 The spin dependence of ${\cal G}_{{\bm k} \sigma}$ turned out to be unimportant (not essential to  AHE, just giving higher-order corrections), which will thus be neglected in the following: ${\cal G}_{{\bm k} \sigma} = {\cal G}_{\bm k} $. 
  Using the identity $ {\rm tr}\left( \sigma _i \sigma _j \sigma _k \right) = 2i \epsilon _{ijk}$, we obtain  
\begin{align}
  \left\langle {\rm T}_{\tau }\left[ j_x \left( \tau \right) j_y \right] \right\rangle 
&= \left( \frac{J}{\hbar \gamma N}\right) ^3 \int d1 \int d2 \int d3 
\nonumber \\
&\hspace{-2.5cm}\quad \times 2i \left\langle {\rm T}_{\tau }\left[ \hat{\bm{m}}\left( 1\right) \cdot \left\{ \hat{\bm{m}}\left( 2\right) \times \hat{\bm{m}}\left( 3\right) \right\} \right] \right\rangle \mathcal{E}_{xy}\left( 1,2,3\right)\,,\,
\end{align}
where $\mathcal{E}_{xy}\left( 1,2,3 \right) $ is the electron part consisting of ${\cal G}$'s, and 
$1 \equiv ({\bm r}_1, \tau_1)$, $\int d1 \equiv \int d\tau_1 \int d{\bm r}_1$, etc. 
 This expression can be regarded as a generalization of the spin-chirality formula of Hall conductivity obtained in Ref.~\cite{Tatara2002} to the situations where the spin texture is allowed to fluctuate. 

\begin{figure}[bt]
\includegraphics[width=1\linewidth]{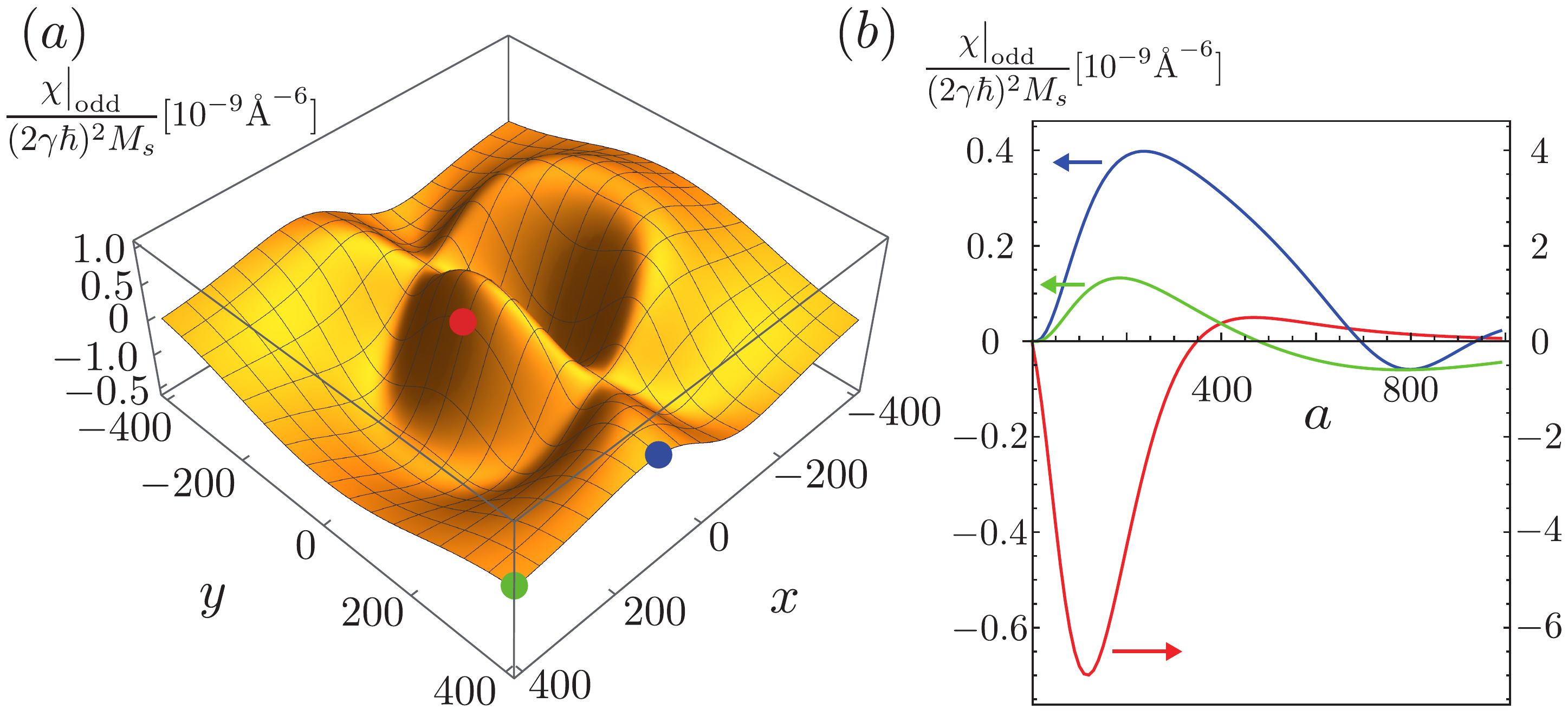}
\caption{\label{fig:chirality}(color online) Chiral spin correlation $\chi |_{\rm odd}$. 
(a) As a function of $\bm{r}_3=(x,y,0)$ for  $\bm{r}_1 =(0,200,0)$, $\bm{r}_2 =(0,-200,0)$ and $\tau _1 = \tau _2 = \tau _3$. 
The unit of length is {\rm \AA}. 
 (b) As a function of $a =\left| \bm{r}_1 -\bm{r}_2 \right| $ at $(x,y) =$ (100,0), (100,400) and (400,400) plotted by red, blue and green lines, respectively.  
 As seen, $\chi |_{\rm odd}$ is odd in $x$ while oscillating in $x,y$ and $a$, consistent with the vanishing spatial average of $\left( \bm{a}\times \bm{b}\right) _z \chi $.}
\end{figure}

In order to have non-vanishing $\sigma_{xy}$, it is necessary that the time-ordered chiral spin correlation
\begin{equation} 
 \chi = \left\langle {\rm T}_{\tau }\left[ \hat{\bm{m}}\left( 1\right) \cdot \left\{ \hat{\bm{m}}\left( 2\right) \times \hat{\bm{m}}\left( 3\right) \right\} \right] \right\rangle 
\end{equation}
has parity-odd components  since $\mathcal{E}_{xy}\left( 1,2,3\right) $ is parity odd. 
 As anticipated, the spin-flip magnon propagators of Eq.~(\ref{Dpm}) serve this purpose. 
 In fact, one can derive from Wick's theorem the parity-odd part of  $\chi$ as 
\begin{align}
\chi \, \big| _{\rm odd} &= i\hbar ^2 \gamma ^2 M_s \Big[ \mathcal{D}_{--}\left( 1,2\right) \mathcal{D}_{++}\left( 1,3\right)  \label{eq:chirality} \\
& - \mathcal{D}_{++}\left( 1,2\right) \mathcal{D}_{--}\left( 1,3\right) + ({\rm cyclic \ perms}) \Big] \,.\nonumber
\end{align}
Here we expanded $\hat m_z$ to the lowest order in the magnetization fluctuation as
$ \hat{m}_z  = M_s - \hat{m}_- \hat{m}_+  /(2M_s)$. 
 More explicitly, this takes the form
\begin{align}\label{eq:chi_2}
 \frac{\chi \, \big| _{\rm odd} }{(2\gamma \hbar) ^2 M_s} 
&=  ({\bm a} \times {\bm b})_z 
    \Bigl[  ({\bm a} \cdot {\bm b})_t F_a F_b +  ({\bm b} \cdot {\bm c})_t F_b F_c 
\nonumber \\ 
& \hskip 18mm 
 + ({\bm c} \cdot {\bm a})_t F_c F_a  \Bigr]  ,   
\end{align}
where $ F_a \equiv \left( a_x \pm ia_y \right) ^{-2}\mathcal{D}_{\pm \pm }\left( \bm{a};\tau _a \right) $
depends only on $a_t  = \sqrt{a_x^2+a_y^2} , a_z$ and $\tau_a = \tau_1-\tau_2$
(i.e. $F_a$ is a scalar with respect to rotation around $z$-axis). 
 Equation (\ref{eq:chi_2}) is an elaboration of Eq.~(\ref{eq:chi_1}).
 Spatial variation of  $\chi \, \big| _{\rm odd} $ is visualized in FIG.~\ref{fig:chirality}. 
We used $\lambda = 0.4 \times 10^{-16} \,  [{\rm m}^2], \hbar \omega _0 = \hbar \omega _M = 10^{-4} \, [{\rm eV}], k_B T = 10^{-2} \, [{\rm eV}] $ as a representative set of parameters.
 The ${\bm q}$-integral in Eq.~(\ref{Dpm}) was evaluated by introducing a cutoff $q_m = 1 \, [{\rm \AA}^{-1}]$. 
Remarkably, there is a definite chiral spin correlation with odd parity once the relative locations (${\bm a}$ and ${\bm b}$) are fixed.
 If averaged over the locations, however, it vanishes because of the factor $({\bm a} \cdot {\bm b})_t$, as mentioned in the introduction.

\begin{figure}[bt]
\includegraphics[width=1\linewidth]{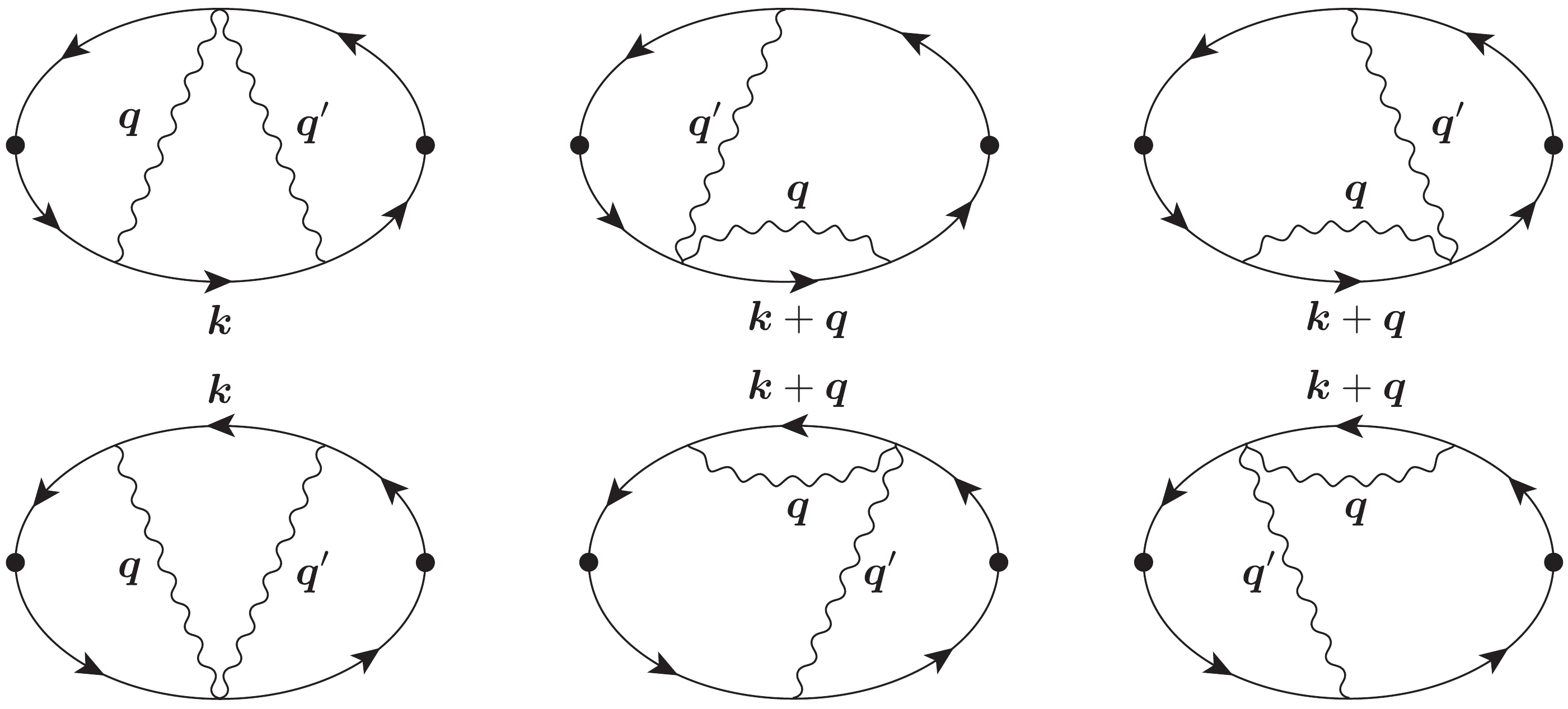}
\caption{\label{fig:relevant}The diagrams contributing to $\sigma _{xy}$ at $J^3$. 
In terms of $\mathcal{E}_{xy}\left( 1,2,3 \right) $, there are only two configurations; those in the top row and those in the bottom. The three variations within each row occur due to the cyclic permutation of magnon arguments in Eq.~(\ref{eq:chirality}). 
 Electron-magnon vertices with one (two) wavy line(s) represent $x$, $y$ components ($z$ component) in $H_{\rm sd}$. 
}
\end{figure}

 The Hall conductivity $\sigma_{xy}$ is obtained by combining the magnon part $\chi$ with the electron part $\mathcal{E}_{xy}\left( 1,2,3\right)$. 
 The resulting diagrams are shown in FIG.~\ref{fig:relevant}, where the solid and wavy lines represent $\mathcal{G}$ and $\mathcal{D}$, respectively. 
 Every term of $\mathcal{E}_{xy}$ is a product of a factor $(\bm{a}\times \bm{b})_z $ (or its equivalent) and functions of $a,b$ and $c$. 
 Since $c= |{\bm a} + {\bm b}|$ contains a factor $(\bm{a}\cdot \bm{b})_t$, the sign change due to $(\bm{a}\cdot \bm{b})_t$ in Eq.~(\ref{eq:chi_2}) upon integration is compensated, resulting in a non-vanishing $\sigma _{xy}$. 
 This is to be compared with the case of Ref.~\cite{Tatara2002}, in which the electron part that is independent of $(\bm{a}\cdot \bm{b})_t$ contributed to $\sigma_{xy}$.

\begin{figure}[tb]
\includegraphics[width=1\linewidth]{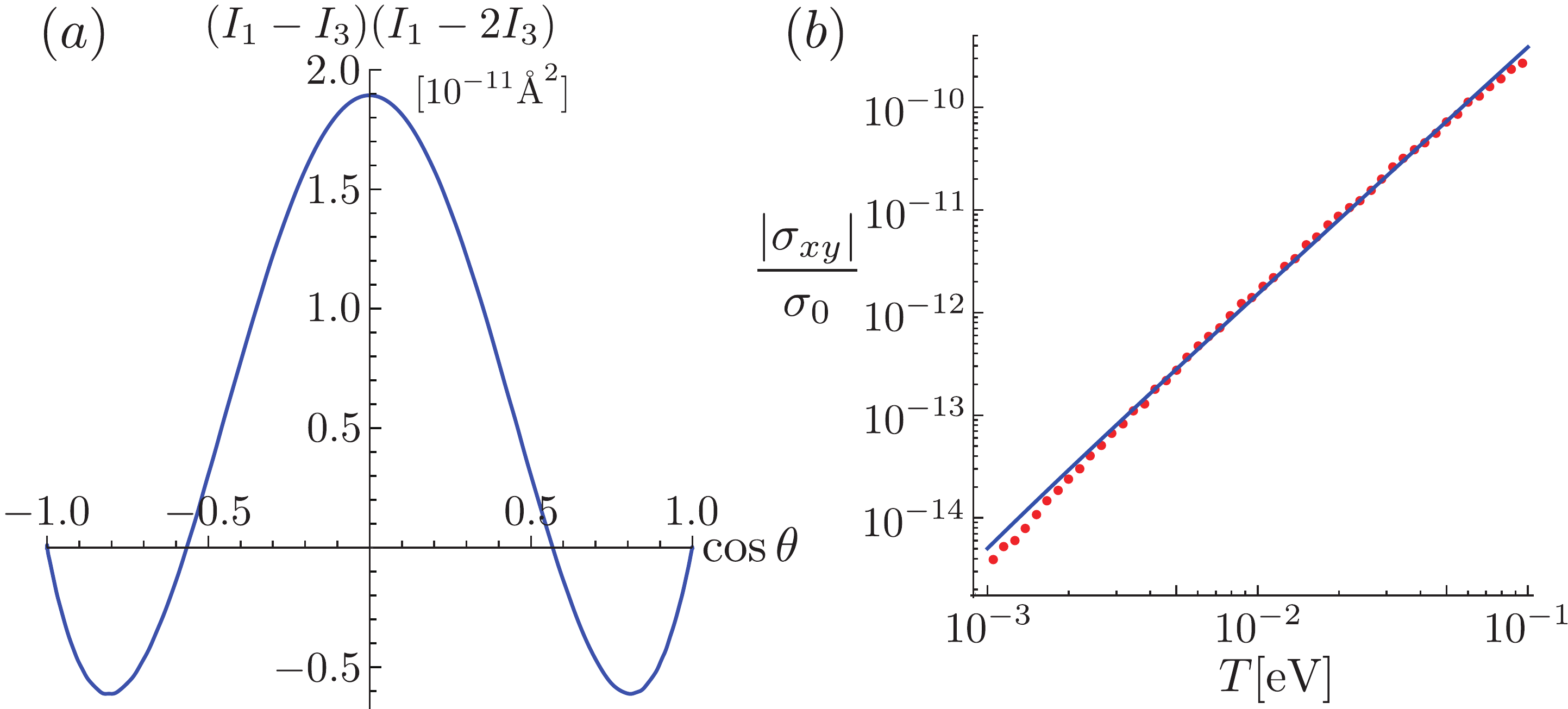}
\caption{\label{fig:results}(color online) (a) $( I_1 -I_3 ) (I_1 - 2I_3 ) $ in Eq.~(\ref{eq:res}) 
for $(k_t, k_z) = k_F (\sin \theta , \cos \theta)$ as a function of $\cos \theta $. 
 Note that $k_t d\theta = -k_F d( \cos \theta )$. 
 (b) Hall conductivity evaluated by Monte Carlo integration (red dots) 
with a fit $|\sigma _{xy}|/\sigma _0 = 6.83 \times 10^{-8}  \times T^{2.29} e^{-0.001/T}$ (blue line). Plain power-law fitting gives  $\sim T^{2.48}$.   
The same parameter set as in FIG.~\ref{fig:chirality} has been used along with $\mu = 0.6\, [{\rm eV}]$. }
\end{figure}

 To evaluate $\sigma_{xy}$ explicitly, it is convenient to work with the Fourier (wave-vector and Matsubara frequency) space. 
 After performing analytic continuation $i\omega_\lambda \rightarrow \hbar \omega +i0$ and taking the limit $\omega \rightarrow 0$, one finds a number of terms containing different combinations of retarded and advanced Green's functions. 
 As we are computing a transport quantity, these terms are naturally classified according to the powers of the electron lifetime, or decay rate $\Gamma_{\bm k} \left( \epsilon \right) $. 
 In the clean limit, the dominant term is that of the lowest power in $\Gamma $ \cite{Nagaosa2010},
originating from 
the two pairs of retarded and advanced electron propagators at both ends of the diagrams with identical momentum arguments.
 This is similar to the so-called extrinsic contribution and reads 
\begin{widetext}
\begin{eqnarray}
 \sigma _{xy} &=& -\frac{2\hbar ^3 e^2 J^3 S V^2}{\pi^3 m^2 N^2}
 \int d\epsilon \int d\epsilon _1  \int d\epsilon _2 \sum_{{\bm k}, {\bm q}, {\bm q}'} 
 A^+_{\bm{q}}A^-_{\bm{q}}A^+_{\bm{q}^{\prime }}A^-_{\bm{q}^{\prime }} \sin [2\left( \phi _{\bm{q}}-\phi _{\bm{q}^{\prime }}\right)] f^{\prime } \left( \epsilon \right)   \nonumber \\
&& \times  \left\{ n\left( \epsilon _1 - \epsilon \right) + f\left( \epsilon _1 \right) \right\} 
                \left\{ n\left( \epsilon _2 -\epsilon \right) + f\left( \epsilon _2 \right) \right\} 
                \left| G_{\bm{k}+\bm{q}^{\prime } }^{\rm R} \left( \epsilon \right) \right| ^2 
                {\rm Im} \, D_{\bm{q}}^{\rm R} \left( \epsilon _1 -\epsilon \right) 
                {\rm Im} \, D_{\bm{q}^{\prime } }^{\rm R} \left( \epsilon _2 -\epsilon \right)
\label{eq:integral}\\
&& \times \left[ \left\{ \left( \bm{k} +\bm{q} \right) \times \left( \bm{k}+\bm{q}^{\prime } \right) \right\}_z 
                      {\rm Im} \, G_{\bm{k}}^{\rm R} \left( \epsilon \right) 
                     \left| G_{\bm{k}+\bm{q}}^{\rm R} \left( \epsilon_1 \right) \right| ^2 
                  - 2 \left( \bm{k}\times \bm{q}^{\prime } \right)_z 
                      \left| G_{\bm{k}}^{\rm R} \left( \epsilon \right) \right| ^2 
                      {\rm Im} \, G_{\bm{k}+\bm{q}}^{\rm R} \left( \epsilon _1 \right)  \right]\,, \nonumber
\end{eqnarray} 
\end{widetext}
where $n(x) $ and $f(x)$ are Bose and Fermi distribution functions, and 
$G^{\rm R}$ and $D^{\rm R}$ are the retarded counterparts of $\mathcal{G}$ and $\mathcal{D}$, respectively. 
 In the square bracket in the last line of Eq. (\ref{eq:integral}), the first term comes from the diagrams in the first column of FIG.~\ref{fig:relevant} while the second term from the sum of the rest. 
  To the lowest order in $\Gamma$, one can replace $ \left| G_{\bm k}^{\rm R} \left( \epsilon \right) \right| ^2$  
by $\pi \delta \left( \epsilon -\xi _{\bm{k}} \right)/ \Gamma_{\bm k} \left( \epsilon \right)  $ and other propagators by their unperturbed form. 
 The products of the distribution functions effectively restrict the energy variables $\epsilon, \epsilon _1 $ and $\epsilon _2$ 
to  $\left[ -k_B T ,k_B T \right] $, much smaller than $\mu$, and the delta functions force $\bm{k}$, $\bm{k}+\bm{q}$ and $\bm{k}+\bm{q}^{\prime }$ to be near the Fermi surface and $\Gamma $'s can be regarded as a constant $\Gamma_F \equiv \Gamma_{k_F}(\epsilon_F)$. 
 Integrating over energies and $\phi _{\bm{k}}$, we arrive at
\begin{eqnarray}
\sigma _{xy} &=& \sigma _0 \frac{\epsilon _F}{k_F} \int_{-\infty}^\infty dk_z  \int_0^\infty dk_t f^{\prime }\left( \xi _{\bm{k}} \right) k_t \label{eq:res} \\
&& \times \left\{ I_1 \left( \bm{k} \right) - I_3 \left( \bm{k} \right) \right\} \left\{ I_1 \left( \bm{k} \right) - 2I_3 \left( \bm{k} \right) \right\}\,,  \nonumber 
\end{eqnarray}
where $\sigma _0 \equiv J^3 S e^2 k_F / (2\pi \hbar \epsilon _F \Gamma_F ^2)$, 
$\bm{k}$ now denotes doublet $(k_t ,k_z )$, $k_t =\sqrt{k_x^2 +k_y^2} $
and the functions $I_n \left( \bm{k}\right) $ are given by
\begin{eqnarray}
 I_n &=&\sum _{\alpha =\pm 1} \frac{\hbar ^2}{2m}\frac{V}{N} \int \frac{d^3 \bm{q}}{(2\pi )^3} 
                                          \frac{\omega _M}{\omega _{\bm{q}} } 
                                          \left[ \frac{(\bm{k}\cdot \bm{q})_t}{q_t k_t} \right]^n q_t 
\label{eq:I} \\
&& \times  \left\{ n\left( \hbar \omega _{\bm{q}} \right) + f\left( \alpha \xi _{\bm{k}} + \hbar \omega _{\bm{q}} \right) \right\} 
       \delta \left( \xi _{\bm{k}+\bm{q}} - \xi _{\bm{k}}-\alpha \hbar \omega _{\bm{q}} \right) . \nonumber
\end{eqnarray}
 We note $\sigma _{xy}$ is negative for $JS>0$. 
 This is seen by approximating $-f^{\prime }$ by delta function; then the integral is essentially the area under the curve in FIG.~\ref{fig:results} (a). 
Temperature dependence has been studied numerically, yielding a roughly power-law dependence $\sigma _{xy} \propto T^p $ with a small, but appreciable, deviation [FIG.~\ref{fig:results} (b)]. The estimated value of $p$ falls in the range $ 2.2 \lesssim p \lesssim 2.5$ depending on the choice of fitting function.

The mechanism of the present AHE can be identified as the skew scattering by the chiral spin correlations of dipolar magnons.
 In fact, by relabelling the wave-vectors in FIG.~\ref{fig:relevant}, the on-shell scattering amplitude of two electrons with momenta $\bm{p}$, $\bm{p}^{\prime }$ at the ends of the diagrams  
going through $\chi \, |_{\rm odd}$ (with an intermediate electron line) 
can be written as proportional to $( \bm{p}\times \bm{p}^{\prime })_z $.

To conclude, we have shown that there is a nonzero contribution to the AHE originating from the interaction with the dipolar magnons in homogeneous ferromagnets. 
 The leading order term occurs at the third order in the {\it s-d} exchange interaction and involves a factor of the chiral spin correlation felt by traversing electrons. 
 The result serves as a generalization of the Hall effect from static spin chirality as well as giving a concrete example in which anomalous Hall current requires neither relativistic spin-orbit interaction nor fixed magnetic texture. 
 In a broader sense, the dipole-dipole interaction acts as a kind of spin-orbit interaction in that it couples the spin and orbital motion of magnons.
%{\color{red} it induces a correlation between the direction of spin and wave-vector for magnons.}
Electrons then inherit the chiral orbital motion through skew scatterings by the quantum/thermal dipolar magnons.
 
Unfortunately 
the predicted Hall conductivity is small because of the dominance of exchange energy over the dipole-dipole term, i.e. $\lambda q^2 \gg 1 $ for typical $q$, which is $\sqrt{k_B T / (\hbar \omega _M \lambda )} $.
 To identify the proposed AHE experimentally, therefore, material should be chosen carefully. First of all, conventional AHE due to spin-orbit interaction must be suppressed in conductors; for example, in some ferromagnetic metals, anomalous Hall coefficients are known to be compensated by blending elements with opposite signs of AHE. 
%Since further numerical studies (not shown) indicate that $|\sigma _{xy}|$ decreases with $\epsilon _F$ as long as $\epsilon _F \gg k_B T$, 
%using a semiconductor magnet or a hybrid film comprising a ferromagnetic insulator layer and a low-carrier conductor layer may be a plausible choice to lower the Fermi energy. 
 Since further numerical studies (not shown) indicate that $|\sigma _{xy}|$ decreases with $\epsilon _F$ as long as $\epsilon _F \gg k_B T$, 
materials with small Fermi energy will be favorable, such as 
 a semiconductor magnet or a hybrid film comprising a ferromagnetic insulator layer and a low-carrier conductor layer.
In many materials, AHE accompanies spin Hall effect which is often used in spintronics, and the present mechanism may open the door to spintronics free from spin-orbit interaction.

 \begin{acknowledgements}
 We would like to thank Shuichi Murakami, Jairo Sinova and Yaroslav Tserkovnyak for helpful discussions.
 This work was performed under the Inter-university Cooperative Research Program of the Institute for Materials Research, Tohokou University (Proposal No.15K0089). 
 It is also supported by World Premier International Research Center Initiative (WPI), MEXT, Japan. 
 K. Y. is supported by the JSPS Grant-in-Aid for Scientific Research No.26$\cdot$1204. 
 E. S. is supported by Grant-in Aid for Scientific Research on Innovative Areas "Nano Spin Conversion Science" (26103005). 
 K. S. is supported by JSPS KAKENHI Grant No.~15K13531.  
 H. K. is supported by JSPS KAKENHI Grant No.~25400339. 

 \end{acknowledgements}

\bibliographystyle{apsrev4-1}

%\bibliography{bibliography_spin2}

\end{document}